\documentstyle[prl,aps,preprint,epsfig]{revtex}

\def\btabl{\begin{table}}   \def\etabl{\end{table}}
\def\bnn{\begin{eqnarray*}}   \def\enn{\end{eqnarray*}}
\def\btabu{\begin{tabular}}   \def\etabu{\end{tabular}}
\def\bec{\begin{displaymath}} \def\eec{\end{displaymath}}
\def\eqref#1{(\ref{#1})}

\def\Journal#1#2#3#4{{#1} {\bf #2}, #3 (#4)}

\def\NPB{{\em Nucl. Phys.} B}
\def\PLB{{\em Phys. Lett.}  B}
\def\PRL{\em Phys. Rev. Lett.}
\def\PRD{{\em Phys. Rev.} D}
\def\ZPC{{\em Z. Phys.} C}
\def\EPJ{{\em Eur. Phys. J} C}

\def\be{\begin{equation}}
\def\ee{\end{equation}}
\def\bea{\begin{eqnarray}}
\def\eea{\end{eqnarray}}

\begin{document}
\preprint{\vbox{\baselineskip=13pt
\rightline{CERN-TH/99-127}
\rightline{hep-ph/9905363}}}

\title{$B_s$ decays: CP asymmetries in left-right models \\ with
spontaneous
CP violation}

\author{JOAQUIM MATIAS~\footnote{
Talk given at XXXIV$^th$ Rencontres de Moriond: Electroweak Interactions
and
Unified Theories, Les Arcs, France, 13-20 Mar 1999.
}}
\vskip -0.5cm
\vspace{-0.5cm}

\address{TH-Division, CERN, Geneva 23, Switzerland }

\maketitle
\begin{abstract}
We present the contributions of new CP phases in CP asymmetries of
two-body neutral
$B_s$ decays coming from a left--right model with spontaneous
CP violation. Large deviations from the Standard Model predictions
can be accommodated in a natural way by this type of models.
The new physics effects on the mixing, width difference and decays
are analysed. In particular, we show how the measurement of the angle
$\gamma$ in electroweak penguin-dominated processes
can be largely affected.
\end{abstract}

\vspace{7.5cm}

\leftline{May 1999}

\newpage
\section{Introduction}

CP violation in B decays and its measurement using CP asymmetries is one
of the major targets of  B factories and of B experiments at hadron
facilities.
The Standard Model (SM) has specific predictions on the size as well as on
the
pattern of CP violation in $B_{d,s}$ meson decays. Since these predictions
can be tested in these experiments, deviations from them
would signal New Physics (NP). This justifies the
big effort that is being done both on the experimental and on
the theoretical side.

Moreover, on the theoretical side, certain CP-violating asymmetries in
neutral B
decays  are particularly clean, i.e.,
free from
hadronic uncertainties or with controlled ones using complementary tools
such as isospin analysis or other symmetries.
In the context of the SM they would allow for a clean extraction of the
CKM phases, while if NP is at work they would be quite sensitive.
Clear signals of NP can be detected by measuring asymmetries
that are predicted to be zero in the SM  or by comparing two asymmetries that measure the same
angle in the SM but that are differently affected by NP.

In the SM the $B_{d,s}$ systems have  been extensively
studied~\cite{burf}.
 There are also a number of
studies of the NP
effects in $B_d$ decays \cite{gross}. However,
the $B_s$ system has received somewhat less attention from the NP
point of view \cite{gros}.
Very fast oscillations of the $B_s$ system require
outstanding experimental sensitivity (not yet achieved)
 to measure  time-dependent asymmetries.
However, due to the large width difference $\Delta \Gamma^{(s)}$, the
$B_s$ system offers new possibilities for testing NP  which do not
exist in the $B_d$ system.

In this talk we will present an example of how NP can affect
the CP asymmetries in two-body $B_s^0$ decays~\cite{bbmr} using a specific
model:
a left--right-symmetric model (LRSM) with spontaneous CP
violation~\cite{lrsm}.
This model has, concerning CP violation, two  very interesting
features. On the one hand, it is a natural extension to CP of the idea
of parity as a spontaneously broken symmetry, with no need for
the Higgs sector to be enlarged. On the other hand, due to the  new
phases that
appear in the model, the LRSM with spontaneous CP violation
is able to accommodate in a natural way large deviations from
the SM predictions,
if they are seen in future experiments.

\section{The model: LRSM with spontaneous CP violation}

This model is a gauge extension of the SM based on the gauge group:
$SU(2)_L\times SU(2)_R \times U(1)_{B-L}$~\cite{lrsm}. It contains, in
addition
to the standard gauge bosons, an extra $W^{\prime \pm}$ and
$Z^{\prime}$.
The two charged gauge bosons mix with a mixing angle $\xi^{\pm}$.

The fermionic sector is organized in $SU(2)$ left and right doublets
with respect to the corresponding gauge group. The Higgs sector
contains a
bidoublet $\phi$ that gives masses to the extra gauge bosons and two
triplets ($\Delta_R$ and $\Delta_L$), one of them with a large $v_R$
vacuum expectation value coupling
primarily to the
$W_R$ and the other that is included to preserve the left--right
symmetry.
The choice of this scalar sector also allows us to give a natural
explanation of the smallness of the neutrino masses.

The gauge-symmetry breaking proceeds in two stages. In a first stage,
the neutral component of $\Delta_R$ acquires a vev $v_R$
and breaks the symmetry group to $SU(2)_L \times U(1)_Y$, breaking also
the parity symmetry. In a second stage the vev's of the bidoublet
$\phi$
\bea
\langle \phi \rangle  =
\pmatrix {\frac{k_1}{\sqrt{2}} & 0 \cr 0 & \frac{k_2}{\sqrt{2}}\cr},
\eea
completely break  the gauge group down to $U(1)_Q$.
If one allows for the possibility of the vev's having phases, it is
easy to
see, using all the freedom in redefining fields, that we have the choice
of two phases at will. For instance, we can take
$k_2=|k_2| e^{i \alpha}$ and $v_R=|v_R| e^{i \eta}$, breaking
 CP  spontaneously. The phase $\alpha$ will be the
relevant one when dealing with CP violation in the quark sector.
Also the neutral component of $\Delta_L$ acquires a vev $v_L$.

Finally, in order to discuss the CP-violation effects, it is useful to
show
explicitly the charged-current lagrangian in the quark mass eigenstate
basis given by
$
{\cal L}_{CC}={g/\sqrt{2}}(W_L^{\dagger \mu} {\bar u}_L K_L \gamma_\mu d_L
+W_R^{\dagger \mu} {\bar u}_R K_R \gamma_\mu d_R) +h.c.$,
where the left and right CKM matrices ($K_L$ and $K_R$ respectively) are
related by $K_{L}=K_{R}^*
$. The details of the implications in the phase structure of the model
of the previous relation between CKM matrices can be found
in~\cite{bbr1}.

\section{CP asymmetries in the $B_{d/s}$ system}

The time-dependent CP asymmetry for the decays that were tagged as pure
$B_q^{0}$ or ${\bar B}_q^0$ into a common CP eigenstate defined by
\bea
a^{(q)}_{CP}(t)\equiv {\Gamma(B_q^0(t) \rightarrow f) -
\Gamma({\bar B_q^0}(t) \rightarrow f) \over \Gamma(B_q^0(t) \rightarrow f)
+\Gamma({\bar B_q^0}(t) \rightarrow f)}\,,
\eea
is given explicitly by
\bea \label{asym}
a^{(q)}_{CP}(t)=-{ \left( |\lambda^{(q)}|^2-1 \right)
{\rm cos} (\Delta M^{(q)}
t) - 2
{\rm Im} \lambda^{(q)} {\rm sin} (\Delta M^{(q)} t)  \over
(1+ |\lambda^{(q)}|^2) {\rm cosh} ({\frac{1}{2} \Delta \Gamma^{(q)} t}) -2
{\rm
Re}
\lambda^{(q)}
{\rm sinh} ({\frac{1}{2} \Delta
\Gamma^{(q)} t })} \,,
\eea
where
\bea \label{lam}
\lambda^{(q)}=\left(
\sqrt{\frac{M_{12}^{(q)*}-\frac{i}{2}\Gamma_{12}^{(q)*}}
{M_{12}^{(q)}-\frac{i}{2}\Gamma_{12}^{(q)}}}\right)\frac{\bar
A_{q}}{A_{q}}\sim
e^{-2i\phi_M^{q}}
\frac{\bar A_{q}}{A_{q}}\,,
\label{lambda}
\eea
with $\Gamma_{12}^{(q)}$ and $M_{12}^{(q)}$ being the off-diagonal terms
of
the $B_0$--${\bar B}_0$ mixing matrix, $\phi_M^{q}$ the weak mixing
phase and $\Delta \Gamma^{(q)}=\Gamma_H^{(q)}-\Gamma_L^{(q)}$ and
$\Delta M^{(q)}=M_H^{(q)}-M_L^{(q)}$ are the differences in  decay
rates and masses  between the physical eigenstates, respectively.
In the previous expression CPT and
$|\Gamma_{12}^{(q)}|\ll |M_{12}^{(q)}|$ have been assumed.
Concerning the width differences while $\Delta \Gamma^{(d)}$ is  very
small
in the SM,
$\Delta \Gamma^{(s)}$ is expected to be large providing us with another
observable, which could be measured using for instance the untagged
$B_s$ rates $\Gamma(B_q^0 \rightarrow f)+\Gamma({\bar B_q^0}
\rightarrow f)$.

\section{ New physics effects}
Finally we will show how the LRSM affects the CP asymmetries described
above. It will give contributions to $\Delta M^{(q)}$, $\Delta
\Gamma^{(q)}$,
$\phi_M^{q}$ and ${\bar A}_q/A_q$~\cite{bbmr,bbr3}.

In order to compute the new contributions to $\Delta M^{(q)}=2
|M_{12}^{(q)}|$ and
$\Delta
\Gamma^{(q)}=2 |\Gamma_{12}^{(q)}| |\cos (2 \xi)|$~\cite{gr} with
$2\xi={\rm arg}(-M_{12}^{(q)}
\Gamma_{12}^{(q)*})$
one needs to compute the dispersive and absorptive parts
of the box diagrams depicted in \cite{bbr2}. While $M_{12}^{(q)}$ gets a
significant contribution from the diagrams with an exchange of one
$W^{\prime \pm}$ and one $W^{\pm}$ or associated Goldstone boson,
$\Gamma_{12}^{(q)}$ is practically keep untouched, since the
new contribution is strongly suppressed by
$\beta=M_W^2/M_{W^{\prime}}^2$.
However, $\xi$ that enters  $\Delta \Gamma^{(q)}$ and also
 the weak
mixing phase $\phi_M \rightarrow \phi_M^{SM}+2 \xi$, can range between 0
and $\pi$ depending on the values of $\alpha, M_{W^{\prime}}$
and the mass
$M_H$ of the two neutral flavour-changing Higgs-bosons whose masses are
taken to
be equal. This implies that the weak mixing phase
can also
range between these
values so that the asymmetry can accommodate large deviations from the
SM relation. Concerning $\Delta \Gamma^{(s)}$, since in
the SM $\cos(2 \xi) \sim 1$ the overall effect of the LRSM can only be
to reduce $\Delta \Gamma^{(s)}$. Indeed if a drastic reduction is
observed experimentally it could be explained quite naturally by this
model~\cite{bbmr}.

New Physics can also induce modifications in the decay amplitudes~\cite{bbmr}.
Mainly, two situations can arise, depending on the type of CP asymmetry
one is looking at:
either there exist CP asymmetries dominated by tree-level processes whose
SM
contribution vanishes or  is small ($\beta^{\prime}$) (i.e. $B_s
\rightarrow \psi \phi, B_s \rightarrow \psi K_s$). All of them would
be affected by the new contribution coming from the $B^0$--${\bar B^0}$
mixing phase and therefore large departures from the expected zero are
possible. However, notice that all of them would be modified in the same
way $\beta^{\prime} \rightarrow \beta^{\prime} + \delta_m$, where
$\delta_m=2 \xi$ stands for the new contribution to the mixing phase.
Or a completely different situation occurs when the CP asymmetries are
dominated by pure QCD penguin decays
 (such as $B_s \rightarrow \phi \phi$ or $B_s \rightarrow {\bar
K} K_s$)
or electroweak penguins ($B_s \rightarrow \eta \pi$, $B_s \rightarrow
\phi \pi,...$). In that case, their decays may receive considerable
contribution
from New
Physics. Moreover, since the
NP contribution could be different for each process,
CP asymmetries that were measuring the same angle no longer do.

In order to illustrate this second case we will analyse, as an example,
the flavour- changing decay $b \rightarrow s {\bar s} s$~\cite{bbmr}. We
will follow four
steps. First, one should write the Hamiltonian due to gluon exchange
describing this decay at the scale $M_W$: ${\cal H}_{eff}=-{G_F \alpha_s
\over \sqrt{2} \pi} V_L^{ts*} V_L^{tb} ({\bar
s}[\Gamma_\mu^{LL}+\Gamma_\mu^{LR}] T^a b)({\bar s} \gamma_\mu T^a s)$,
where $\Gamma_\mu^{LL}$ is the SM contribution and $\Gamma_\mu^{LR}=2 i
{{m_b} \over q^2} \tilde E_0^{\prime}(x)[A^{tb} \sigma_{\mu\nu} q^{\nu}
P_R+ A^{ts *} \sigma_{\mu \nu} q^{\nu} P_L]$
 is the new contribution
induced by
$W$ exchange via the right handed current,
 $P_L$ and $P_R$ are the
left and right projectors and $A_{tb}=\xi^{+} {m_t/m_b} e^{i \sigma_1}$
and
$A_{ts}=\xi^{\pm} {m_t/m_b} e^{i \sigma_2}$. This new contribution is
suppressed by the mixing angle $\xi^{\pm}<0.01$, but enhanced by
$m_t/m_b
\sim 60$ and the numerical value of ${\tilde E^{\prime}_0}(x) \sim 4
E^{\prime}_0(x)$.  This enhancement overcomes completely the suppression
due to the mixing angle.
Moreover two phases $\sigma_1$ and $\sigma_2$, functions of
$\alpha$ and the signs of the masses of the quarks, appear allowing for
potential big effects in the asymmetries.
The second step consists in calculating the LO QCD corrections, using
an
OPE to integrate out the top and calculate the Wilson coefficients
$C_i$.
In the basis used to describe this decay NP effects
will enter only into the photonic and gluonic magnetic operators with the
left and right structures ($C_7^\gamma,C_8^G,C_7^{\gamma \prime}$ and
$C_8^{G \prime}$). The third step consists in running down the Wilson
coefficients from the $M_W$ scale to $m_b$. Finally, the last step
is to compute the hadronic matrix elements using some
approximation (we used factorization). It was found that the magnetic
contributions are absorbed into penguin contributions by redefining the
Wilson coefficients.
One can now apply this procedure to the evaluation of processes such as
$B_s
\rightarrow \phi \phi$~\cite{bbmr}. This process is dominated by QCD
penguins and it
receives $30 \%$ contribution from EW penguins. Its asymmetry, which is
expected to be zero in the SM, is largely affected by LRSM with
spontaneous CP violation and can be as large as $0.85$ depending on the
values of the new phases $\sigma_1$ and $\sigma_2$. Another example is
$B_s \rightarrow \eta \rho^0$. In that case
the current--current contribution is CKM-suppressed and the EW penguins
dominate. This process was proposed in~\cite{fleischer2} to measure
$\gamma$.
While the structure of the amplitude in the SM is
$A(B_s\to\eta\rho^0)=A_{CC}e^{-i\gamma}+A_{EW},$ where $A_{CC}$ and
$A_{EW}$ are the current--current and EW penguin contributions,
respectively, if NP
is present an extra piece $A_{NP}e^{-i\phi}$ should be added for each
new phase, where
$A_{NP}$ is the magnitude of the new contribution and $\phi$ its phase.
Then, in the presence of NP, the asymmetry does not measure
anymore $\sin{\gamma}$ but $\sin{\gamma}+z \sin{\phi}$ (for one
extra phase), where $z$ is
defined by
 $z=A_{NP}/A_{CC}$. In the case
of a LRSM with spontaneous CP violation~\cite{bbmr} $z$
can be of order 1 and the two new phases, $\sigma_1$ and
$\sigma_2$, distort completely the measurement of $\gamma$. If the
extracted value of $\gamma$ from a second process differs, that would
signal NP.

In conclusion, we have shown that if large departures from the SM are
found in the width and mass difference of the $B_s^0$ system they can be
accommodated by a model
of LRSM with spontaneous CP violation. Also important effects can be
induced in the asymmetries of the decays that are predicted to be zero
in the SM such as $B_s \rightarrow \phi
\phi$ or in EW penguin-dominated decays such as $B_s^0\to
\eta^{(')}\rho^0,\,\phi\rho^0$.

\section*{Acknowledgements}

J.M. acknowledges the financial support from a Marie Curie EC Grant
(TMR-ERBFMBICT 972147).

\end{document}